\documentclass[12pt]{article}
\usepackage{amsfonts}
\usepackage{amsmath}
\usepackage{amsthm}

\advance \topmargin by -\headheight
\advance \topmargin by -\headsep
     
\evensidemargin \oddsidemargin
\newcommand{\chii}{\raise.5ex\hbox{$\chi$}}
\newcommand{\Z}{{\mathbb Z}}
\newcommand{\R}{{\mathbb R}}

\newcommand{\C}{{\mathbb C}}

\newcommand{\noi}{\vspace{12pt}\noindent}

\newcommand{\ie}{{${ i.e.\ }$}}
\newcommand{\eg}{{${ e.g.\ }$}}
\newcommand{\ea}{{${ et~al.\ }$}}
 
\def\lab#1{\label{eq:#1}}
\def\rf#1{\eqref{eq:#1}}
\newcommand{\eq}[1]{{eq.~\rf{#1}}}
\newcommand{\rfs}[2]{{\rf{#1} and \rf{#2}}}
\newcommand{\eqs}[2]{{eqs.\ \rf{#1} and \rf{#2}}}
\newcommand{\Ref}[1]{{Ref.~\cite{#1}}}

\newcommand{\equi}[1]{\stackrel{{#1}}{=}}

\newcommand{\ct}[1]{\cite{#1}}
\newcommand{\bi}[1]{\bibitem{#1}}

\def\br{\begin{eqnarray}}
\def\er{\end{eqnarray}}
\def\be{\begin{equation}}
\def\ee{\end{equation}}
\def\ba{\be\begin{array}{c}}
\def\ea{\end{array}\ee}
\def\bc{\begin{center}}
\def\ec{\end{center}}

\def\({\left(}
\def\){\right)}

\newcommand{\Ad}{{\cal A}_{\Delta}}

\numberwithin{equation}{section}

\begin{document}
\begin{flushright}
\today
\end{flushright}

\begin{center}
{\large\bf Compatible Poisson Structures}
\end{center}
\begin{center}
{\large\bf of Toda Type Discrete Hierarchy}
\end{center}
\begin{center}
{\sc Henrik Aratyn\footnote{E-mail:~{\tt aratyn@uic.edu}} 
and Klaus Bering\footnote{E-mail:~{\tt bering@uic.edu}}}

\par \vskip .1in \noindent
Department of Physics \\
University of Illinois at Chicago\\
845 W.~Taylor St.\\
Chicago, Illinois 60607-7059\\
\par \vskip .3in

\end{center}

\begin{abstract}
An algebra isomorphism between algebras of matrices and difference operators
is used to investigate the discrete integrable hierarchy. We find local and 
non-local families of $R$-matrix solutions to the modified Yang-Baxter 
equation. The three $R$-theoretic Poisson structures and the Suris quadratic 
bracket are derived. The resulting family of bi-Poisson structures include a
seminal discrete bi-Poisson structure of Kupershmidt at a special value. 
\end{abstract}

\noi

\noi
PACS number(s): 02.10.Sp, 02.10.Tq, 03.20.+i, 04.60.Nc, 11.30.-j, 11.30.Na.

\section{\sf Introduction: A Matrix Formulation of the Lattice Hierarchy}
\label{section:lmf}

\noi
A popular framework for dealing with infinite lattice systems is the 
differential-difference calculus \ct{asterisque} based on a Lax difference 
operator
\be
L  ~=~ \sum_{k \in \Z}  u_k(x) \Delta^k~,  \lab{ourlax}
\ee
where $\Delta$ is the translation operator $\Delta^k f(x)=f(x\!+\!k)\Delta^k$, 
or equivalently $[\Delta,x]=\Delta$. In these cases, the dynamics of the 
field $u_k(x)$ are governed by a Lax equation  
\be
\frac{\partial L}{\partial t_n} \, = \,[ {\cal P}_{+}(L^n),~ L]
~,~~~~~~~~~~~~~~~~~~~~~n \ge 0~,
\lab{leom0}
\ee
where ${\cal P}_{+}$ is a projection-like operator, whose detailed form is 
discussed in Section~\ref{subsection:lf}. It is one of the hallmarks of the 
discrete hierarchy that the dynamical field $u_k(x)$ only interacts with 
itself in points that belong to the same (affine) integer lattice $\Z + x$. 
Therefore, if one assumes that the Hamiltonians ${\cal P}_{+}(L^n)$ do not 
have explicit $x$-dependence, one may ignore the fractional part of the space
coordinate $x$, as it only labels isomorphic non-interacting substructures of
the discrete hierarchy. In other words, it is legitimate to consider the 
space variable $x$ to live on the set $\Z$ of integer numbers rather than the
whole continuous space $\R$. This trivial fact reduces the discrete hierarchy
to an infinite, but countable, matrix problem. The matrix picture becomes even
clearer if one writes the dynamical field $u_k(x)$ as a matrix $u_{ij}$, 
$i,j \in \Z$:
\be
  u_k(x) ~=~ u_{x,x+k} ~=~ u(x,x\!+\!k)       \quad {\rm or} \quad
  u_{ij}~=~u_{j-i}(i)~.
\lab{link}
\ee
It is convenient to call $u_k(x)\equiv u_{x,x+k}$ a {\em link} from a lattice 
point $x\in\Z$ to a lattice point $x+k\in\Z$, and to call the integer
$k\in\Z$ the (signed) {\em length} of the link. 
Translating the difference operator $L$ into a matrix $u=(u_{ij})$
turns out to be very fruitful, partially because this 
provides a better geometric and algebraic understanding. {}For illustrative 
purposes, consider the seminal bi-Poisson structure of \Ref{asterisque}, 
here extended to both positive and negative link lengths
\be
\begin{split}
 \{  u_{n}(x) ,  u_{m}(y) \}_{K1} 
~=~&\frac{1}{2}[\varepsilon(n\!-\!\frac{1}{2})
+\varepsilon(m\!-\!\frac{1}{2}) ]\left[u_{n+m}(x)~\delta_{x+n,y}
-u_{n+m}(y)~\delta_{x,y+m}\right]~, \\ 
 \{  u_{n}(x) ,  u_{m}(y) \}_{K2}
~=~&\frac{1}{2}\sum_{k\in\Z} \left[\varepsilon(k)
+\varepsilon(k\!+\!m\!-\!n)\right]
u_{n-k}(y)~u_{m+k}(x)~\delta_{x+k,y} \\
&+ \frac{1}{2}u_{n}(x)~u_{m}(y)\sum_{k\in\Z} [
\varepsilon(k)+\varepsilon(k\!+\!m\!-\!n) \\
&-\varepsilon(k\!+\!m\!-\!\frac{1}{2})-
\varepsilon(k\!-\!n\!+\!\frac{1}{2})]~\delta_{x+k,y}
~,~~~~~~~~~~~~~~~~~~~n,m\in\Z~,
\end{split}
\ee
and where $\varepsilon$ denotes the sign function. Notice that the 
Kupershmidt Poisson brackets  $\{ \cdot ,\cdot \}_{K1}$ and 
$\{ \cdot ,\cdot \}_{K2}$ are non-local in the $x$ and $y$ coordinates.
It is quite elaborate to verify by brute force that $\{ \cdot ,\cdot \}_{K1}$ 
and $\{ \cdot ,\cdot \}_{K2}$ are a bi-Poisson structure, not to mention 
identifying allowed deformations of this. We shall see that the non-locality
is an artifact of the difference operator language, and that the matrix 
picture, combined with classical $R$-matrix theory, provides an elegant and 
effective formalism to deal with such bi-Poisson structures.

\noi
The outline of the paper is as follows. We continue this 
Section~\ref{section:lmf} with describing the connection between infinite 
matrices and difference operators. This provides a useful connection 
between the Lax formulations of the two pictures. 
In Section~\ref{section:rmatrix}, we revisit classical $R$-matrix theory
for the matrix formulation, where in particular the conditions for 
integrability are given in detail. We provide local and non-local 
classes of solutions to these conditions, that are useful for model building.
In Section~\ref{section:pbrackets}, the Lax equations of motion are recast 
as the Hamiltonian equations of motion by finding viable Poisson brackets, 
that gives rise to bi-Poisson structures. It is natural in this connection 
to briefly review the three $R$-theoretic Poisson brackets and the Suris 
quadratic bracket construction \ct{suris:93}. As a new result, we show that 
the Suris bracket $\{\cdot ,\cdot \}_{S2}$ decomposes into the second 
$R$-theoretic bracket $\{\cdot ,\cdot \}_{R2}$ and a piece 
$\{\cdot ,\cdot \}_{\Omega2}$, that plays no role for on-shell dynamics. 
Our main results for the bracket structure are given in 
\eq{r1bracket}-\rf{r1bracketdiff} and \eq{s2bracket0}-\rf{s2bracketdiff}. 
In Appendix~\ref{section:appkuper}, we translate our Poisson bracket results
into the notation of Kupershmidt \ct{asterisque} for comparison. {}From a 
$R$-theoretic perspective, it is remarkable that already the original 
discrete hierarchy of Kupershmidt induces the generalized construction 
of Suris. {}Finally in Appendix~\ref{section:app22}, we give a $2 \times 2$ 
dimensional example that illustrates aspects of the Suris theory.

\subsection{\sf An Algebra Isomorphism}
\label{subsection:aai}

\noi
We consider an algebra isomorphism from the associative algebra ${\cal A}$ 
of infinite-dimensional matrices (to be specified below) to a certain 
sub-algebra $\Ad$ of difference operators,
\be 
{\cal A}~\ni~u ~~~~~~\mapsto~~~~~~L 
~=~ \sum_{k\in \Z}  u_{x,x+k} \Delta^k~\in~\Ad~.
\lab{algimbed}
\ee
The algebra isomorphism maps matrix multiplication into composition of 
difference operators:
\be
\sum_{k\in\Z}  (uv)_{x,x+k} \Delta^k 
~=~(\sum_{k_1\in\Z}  u_{x,x+k_1} \Delta^{k_1})~ 
 (\sum_{k_2\in\Z}  v_{x,x+k_2} \Delta^{k_2})
~,~~~~~~~~~~~~~~u,v \in {\cal A}~.
\lab{algiso}
\ee
Geometrically, it is a useful fact that the total length $k=k_1+k_2$ of links
is preserved under the matrix multiplication/composition.

\noi
As an easy application, the matrix description provides a pictorial 
understanding for the coefficient functions $(L^n)_k\, (x)$ of the higher 
powers of the Lax operator 
\be
 L^n ~=~ \sum_{k\in\Z}
 \sum_{\substack{k_1, \ldots, k_n\in\Z \\ k_1 + \ldots + k_n=k}}
u_{x,x+k_1} \Delta^{k_1} \cdots  u_{x,x+k_n} \Delta^{k_n}
~=~ 
\sum_{k\in\Z} (L^n)_{k}(x)\, \Delta^k~,~~~~~n \ge 0~.
\lab{lexpand}
\ee
They are
\be
(L^n)_k\, (x) \,= \,
\sum_{\substack{k_1, \ldots, k_n\in\Z \\ k_1 + \ldots + k_n=k}}
\prod_{r=1}^n u \left( x+\sum_{i=1}^{r-1} k_i, x+\sum_{i=1}^{r} k_i \right)
 \,= \, (u^n)_{x,x+k}~,
\ee
or in words, the $k$'th coefficient of the $n$'th power of the Lax operator 
corresponds to $n$ consecutive links with the two free ends a distance 
$k$ apart. Therefore,
\be
 L^n ~=~ \sum_{k\in\Z} (u^n)_{x,x+k}\,  \Delta^k
~,~~~~~~~~~~~~~~~~~~~~~~n \ge 0~.
\lab{lnpower}
\ee
A difference operator, that has all the $x$'s appearing to the left of all 
$\Delta$'s, is called a {\em normal ordered} difference operator.\footnote{
\Ref{asterisque} uses {\em anti-normal ordering}, with all the $\Delta$'s 
appearing to the left of all $x$'s.} 

\noi
Let us now define the matrix algebra ${\cal A}$ itself. {}For simplicity, 
let ${\cal A}$ be the algebra of matrices with only finitely many non-zero 
entries
\be
 {\cal A} \,:= \,  \{ u=(u_{ij}) | u_{ij}=0 ~{\rm for}~ |i|+|j| \gg 0 \} ~.
\ee 
This choice ensures that the matrix multiplication and the matrix trace 
(${\rm tr}$) are well-defined operations.\footnote{\label{footnotalginv}
If one forgets the associative matrix multiplication structure, but keeps 
the Lie commutator operation, ${\cal A}$ is often referred to as the Lie 
algebra ${\rm gl}(\infty)$ in the mathematical literature. Also we stress 
that the above ${\cal A}$ has no identity matrix and no invertible 
matrices in this infinite dimensional case. At a few places in the paper we 
refer to invertible matrices and to make rigorous sense of this, ${\cal A}$ 
should be chosen as a Banach algebra with an algebra norm. It is out of scope 
to provide details here.}

\subsection{\sf Lax Formulation}
\label{subsection:lf}

\noi
In this Section, we return to the Lax equation \rf{leom0} and implement the
corresponding matrix formulation, which appears naturally in the theory of 
classical $R$-matrices (cf.\ the next Section~\ref{section:rmatrix}). The Lax 
equation for $L$ can be written in several ways 
\be
\frac{\partial L}{\partial t_n} 
\, = \,[ {\cal P}_{+}(L^n),~ L] 
~=~[ L,~  {\cal P}_{-}(L^n)]
~=~\frac{1}{2}[ {\cal R}(L^n),~ L]
~,~~~~~~L\in \Ad~,~~~~~n \ge 0~,
\lab{leom}
\ee
for the operators ${\cal P}_{\pm}$ and  ${\cal R}$ related through
\be
{\cal P}_{+}+{\cal P}_{-}~=~{\bf 1} 
 \quad {\rm and} \quad {\cal R}~=~ {\cal P}_{+}-{\cal P}_{-}~.
\lab{tildeppr}
\ee
Although we are going to consider different examples of the operator triples
$({\cal P}_{+},{\cal P}_{-},{\cal R})$, we will always assume that the three
operators ${\cal P}_{+}$, ${\cal P}_{-}$ and ${\cal R}$ are interlocked via
the above two relations \rf{tildeppr}. Hence it is always enough to specify 
one of them. As an example, consider Hamiltonians of the form
\be
{\cal P}_{+} (L^n)\, := \, (L^n)_{\ge 0} + \frac{\nu-1}{2}(L^n)_{0}
~,~~~~~~~~~~~~~~~~~~~n \ge 0~,
\lab{nuham}
\ee
where $\nu$ is a constant parameter \ct{oevel:96}, which is related to a 
choice of operator ordering prescription. The choice $\nu=1$ leads to the 
standard hierarchy with the Hamiltonians given by $(L^n)_{\ge 0}$, while 
$\nu=-1$ leads to the so-called modified hierarchy with the Hamiltonians 
$(L^n)_{ \ge 1}$. 

\noi
To implement the matrix program, one seeks the matrix counterparts 
$R, P_{\pm}: {\cal A} \to {\cal A}$ of the operators
${\cal R}, {\cal P}_{\pm}: \Ad \to \Ad$. They satisfy
\be
P_{+}+P_{-}~=~{\bf 1} 
 \quad {\rm and} \quad R~=~ P_{+}-P_{-}
\lab{ppr}
\ee
as well. Inspired by \eq{lnpower}, we claim that the sought-for identification 
is provided by
\be
\begin{split}
{\cal R}(L^n) ~=~& \sum_{k\in\Z} (R(u^n))_{x,x+k} \Delta^k~, \\
{\cal P}_{\pm}(L^n) ~=~& \sum_{k\in\Z} (P_{\pm}(u^n))_{x,x+k} \Delta^k  
~,~~~~~~~~~~~~~~~~~~~~~~n \ge 0~.
\end{split} 
\lab{idridp}
\ee
It may look discouraging that (for instance) the $P_{+}$ operator does not act 
on the $\Delta$-part at all, but only on the $u$-part, as one usually counts
the power $k$ in the $\Delta^k$-factor to determine the action of 
${\cal P}_{+}$ (cf.\ \eq{nuham}). However, one should recall that $k$ is also
available in the $u$-part as a link length. 

\noi
One can now lift the Lax equation \rf{leom} to the corresponding matrix 
algebra ${\cal A}$: 
\be
 \frac{\partial u}{\partial t_n}~=~[P_{+}(u^n),u]~=~[u,P_{-}(u^n)]
~=~\frac{1}{2}[R(u^n),u]~,~~~~~~u\in {\cal A}~,~~~~~n \ge 0~.
\lab{ueom}
\ee
In fact, the equivalence of the Lax \eq{leom} for the difference operator 
$L \in \Ad$ and the Lax \eq{ueom} for the matrix $u\in {\cal A}$ follows 
straightforwardly from the algebra isomorphism \rf{algiso} and the 
prescription of \eq{idridp}. 

\noi
Let us work out the corresponding matrix maps 
$R, P_{\pm}: {\cal A} \to {\cal A}$ of the example \eq{nuham} given above. 
To this end, we need to introduce some notation. Let $e_{ij} \in {\cal A}$ 
denote an elementary matrix such that 
$(e_{ij})_{kl} = \delta_{i,k} \delta_{j,l}$. As is well-known, the $e_{ij}$'s 
constitute a standard basis for the matrix algebra ${\cal A}$, and 
hence a generic algebra element $u$ can be decomposed as 
$u=\sum_{i,j\in\Z}u_{ij}e_{ij} \in {\cal A}$. Similarly, one only have to 
determine the linear maps $R, P_{\pm}: {\cal A} \to {\cal A}$ on the basis
$e_{ij}$. The conversion prescription of \eq{idridp} is satisfied if one let
\be
\begin{split}
R(e_{ij})~=~&\varepsilon_{\nu}(j\!-\!i)~e_{ij}~,  \\
P_{+}(e_{ij})~=~&\theta_{\nu}(j\!-\!i)~e_{ij}~,  \\
P_{-}(e_{ij})~=~&\theta_{-\nu}(i\!-\!j)~e_{ij}~, 
\end{split}
\lab{glob}
\ee
where we have defined a sign function $\varepsilon_{\nu}(x)$ as
\be
\varepsilon_{\nu}(x):=
 \begin{cases}  
1  &{\rm for}~x>0  \\
\nu&{\rm for}~x=0\\
-1 &{\rm for}~x<0~,
\end{cases}
\lab{epsilonu}
\ee
and a corresponding step function
\be 
\theta_{\nu}(x)
~:=~\frac{1}{2}[1+\varepsilon_{\nu}(x)]~=~
 \begin{cases}  
1  &{\rm for}~x>0  \\
\frac{1+\nu}{2}&{\rm for}~x=0 \\
0 &{\rm for}~x<0~.
\end{cases}
\ee

\section{\sf $R$-Matrix Formalism}
\label{section:rmatrix}

\noi
The classical $R$-matrix theory provides a universal method for constructing 
three compatible Poisson structures and infinitely many commuting charges 
for a wide class of integrable models \ct{sts:83}.\footnote{
See also \eg \ct{suris:93,oevel:89,li:89,Aratyn:1992ba,harnad:93,blaszak:97}.
In \ct{sorin:03} an extension of the $R$-formalism to the fermionic
Toda model is given.} 
A classical $R$-matrix is by definition a linear map $R:{\cal A}\to {\cal A}$
such that the $R$ bracket
\begin{subequations}
\br
[u,v]_R&:=&\frac{1}{2}[R(u),v]+\frac{1}{2}[u,R(v)]  \\
&=& [P_{+}(u),P_{+}(v)]-[P_{-}(u),P_{-}(v)]
\er
\end{subequations}
is a Lie-bracket, \ie it satisfies the Jacobi identity. A sufficient 
condition for the Jacobi identity is provided by the modified Yang-Baxter 
equation ${\rm YB}_{\alpha}(u,v)=0$, where the modified Yang-Baxter operator 
is given by
\be
{\rm YB}_{\alpha}(u,v)~:=~[R(u),R(v)]-2R[u,v]_R+\alpha [u,v]~.
\ee
As we already have seen in the previous Section, it is convenient to define 
projection-like operators $P_{\pm}:{\cal A} \to {\cal A}$:
\be
P_{\pm}~:=~\frac{1}{2}({\bf 1} \pm R)
~,~~~~~~~~~~~~~~~~~~~~R~=~P_{+}~-~P_{-}~.
\ee
We emphasize that in general ${\rm Im}(P_{+})$ and ${\rm Im}(P_{-})$ do
{\em not} form a direct sum, and $P_{\pm}$ are {\em not} necessarily
idempotent operators.

\noi
It is instructive to see how the integrable model arises. Consider an 
{\em abelian} subalgebra  ${\cal A}_0 \subseteq {\cal A}$. (Usually we simply 
consider the infinite hierarchy 
${\cal A}_0 \supseteq \{\, u^n \, | \, n\!=\!1,2,\ldots \}$ 
generated by a single algebra element $u \in {\cal A}$.) The dynamical flow 
\be
\delta_v u~=~[P_{+}(v),u]~=~[u,P_{-}(v)]~=~\frac{1}{2}[R(v),u]
~,~~~~~~~~~~u,v \in {\cal A}_0~,
\lab{nflowu}
\ee 
is generated by a Hamiltonian $P_{+}(v)$. After some straightforward 
algebra, the commutator of two flows reads 
\be
[\,\delta_w \,,\,\delta_v\,]u~=~ \frac{1}{4}[N(v,w),u]
~,~~~~~~~~~~~~~~~~~~~~~~~~~u,v,w \in {\cal A}_0~,
\lab{flowcom}
\ee
where we have defined the Nijenhuis tensor \ct{das:89}
\begin{subequations}
\br
  \frac{1}{4} N(u,v)&=&N_{+}(u,v)+ N_{-}(u,v) \lab{nijenhuisn} \\
&=&[P_{\pm}(u),P_{\pm}(v)] \mp P_{\pm}[u,v]_R~,\lab{nijenhuisp}
\er
\end{subequations}
and the chiral Nijenhuis tensors
\be
   N_{\pm}(u,v)~:=~ P_{\mp}[P_{\pm}(u),P_{\pm}(v)]~.
\lab{chiNijenhuis}
\ee
The Nijenhuis tensor is equal to the modified Yang-Baxter operator
\be
    N(u,v)~=~[u,v]+[R(u),R(v)]-2R[u,v]_R~=~{\rm YB}_1(u,v)~.
\ee
The flows $\delta_n$, $n\ge 0$, commute for an integrable system. {}From 
\eq{flowcom}, a sufficient integrability condition is provided by the
modified Yang-Baxter equation 
\be
      {\rm YB}_1(u,v)~\equiv~N(u,v)~=~0~, \lab{suffyb1}
\ee
which our examples in Sections~\ref{subsection:loc}-\ref{subsection:nloc}
will satisfy. More generally, integrability is guaranteed if there exists 
a linear operator $B:{\cal A} \to {\cal A}$, such that
\be
       N(u,v)~=~B[u,v]~,
\ee 
as can easily be checked from \eq{flowcom}. When we assume a vanishing
Nijenhuis tensor $N\!=\!0$, it follows from \eq{nijenhuisp}, that the
$\pm P_{\pm}$ operators are Lie-algebra homomorphisms
$({\cal A},[\cdot,\cdot]_R) \to ({\cal A},[\cdot,\cdot])$, and in particular
that the images ${\rm Im}(P_{\pm})$ are two Lie sub-algebras:
\be
[{\rm Im}(P_{\pm}),{\rm Im}(P_{\pm})]~\subseteq~{\rm Im}(P_{\pm})~.
\ee
The vanishing of the chiral Nijenhuis tensors 
\be
     N_{+}~=~0~~\quad \mbox{and} \quad~~N_{-}~=~0 \lab{npmwedge}
\ee
implies the vanishing of the Nijenhuis tensor $N={\rm YB}_1=0$
(cf.\ \eq{nijenhuisn}). The opposite statement is not true. The local and 
non-local examples in the next 
Sections~\ref{subsection:loc}-\ref{subsection:nloc} will meet the stronger
condition \rf{npmwedge}.

\noi
Both the vanishing Nijenhuis tensor condition \rf{suffyb1} and the vanishing 
chiral Nijenhuis tensor condition \rf{npmwedge} are stable under conjugation 
of the $R$-matrix $R^{\prime}(u)=a R(a^{-1} u a)a^{-1}$ with an invertible 
algebra element $a$. The conjugation procedure can be used to 
generate new $R$-solutions from old $R$-solutions, although we will not 
pursuit this here (cf.\ footnote \ref{footnotalginv}).

\subsection{\sf A Class of Local R-matrix Solutions}
\label{subsection:loc}

\noi
Here we propose a class of local solutions $R, P_{\pm}: {\cal A}\to {\cal A}$ 
to the condition \eq{npmwedge} that is parametrized by an arbitrary function 
$\nu:\Z \to \C$ and that generalizes \rf{glob}. It is given by
\be
\begin{split}
R(e_{ij})~=~&{\cal E}_{\nu}(j,i)~e_{ij}   \\
P_{+}(e_{ij})~=~&\Theta_{\nu}(j,i)~e_{ij}   \\
P_{-}(e_{ij})~=~&\Theta_{-\nu}(i,j)~e_{ij}~,
\end{split} 
\lab{loc}
\ee
where we have defined a generalized sign function (involving now 
$\nu\!=\!\nu(x)$ being a local function)
\be
{\cal E}_{\nu}(x,y)~:=~
 \begin{cases}  
1     &{\rm for}~x>y  \\
\nu(x)&{\rm for}~x=y  \\
-1    &{\rm for}~x<y
\end{cases}
\ee
and a corresponding step function
\be \Theta_{\nu}(x,y)
~:=~\frac{1}{2}[1+{\cal E}_{\nu}(x,y)]~=~
 \begin{cases}  
1  &{\rm for}~x>y  \\
\frac{1+\nu(x)}{2}&{\rm for}~x=y \\
0 &{\rm for}~x<y~.
\end{cases}
\ee
The choice of $\nu:\Z \to \C$ is related to a ($x$-local) choice of 
operator ordering prescription. Here we refer to a $R$-solution as being 
local if the standard basis $e_{ij}$ diagonalizes the $R$-matrix. 
(We emphasize that a local solution usually becomes non-local in terms of 
the difference operator fields $u_{k}(x)$.) We prove in the next
Section~\ref{subsection:nloc} that the chiral Nijenhuis tensors vanish 
(cf.~\eq{npmwedge}), so that $R$ satisfies the modified Yang-Baxter 
equation ${\rm YB}_1(R)=0$.

\noi
The operators  $P_{+}$ and $P_{-}$ from \eq{loc} ``project'' onto (weakly) 
upper or (weakly) lower triangular matrices, respectively, but they may share
diagonal matrices unless $\nu (x)= \pm 1$ :
\be
 {\rm Im}(P_{+}) \cap {\rm Im}(P_{-}) ~\subseteq~ \{0\} 
~~~~~~\Leftrightarrow~~~~~~ {\rm Im}(\nu) ~\subseteq~ \{\pm 1\}   ~.
\lab{ldirectsum}
\ee
Thus in general, the sub-algebras ${\rm Im}(P_{+})$ and ${\rm Im}(P_{-})$ do 
{\em not} form a direct sum. Similarly, the operators $P_{+}$ and $P_{-}$ 
are idempotent ($P_{\pm}^2=P_{\pm}$) if and only if the $R$-matrix is an 
involution ($R^2={\bf 1}$), which holds precisely when $\nu (x)= \pm 1$, as 
expressed by
\be
4~P_{+}~P_{-}~=~{\bf 1}-R^2 ~=~(1\!-\!\nu^2)~\delta~,
\ee
and where $\delta:{\cal A} \to {\cal A}$ projects onto the diagonal matrices
\be
\delta(e_{ij})~:=~\delta_{i,j}~e_{ij}~,~~~~~~~~~
\delta^2~=~\delta~,~~~~~~~~~
R~\delta~=~\nu~\delta~=~\delta~R~.
\ee

\subsection{\sf A Class of Non-Local $R$-Matrix Solutions}
\label{subsection:nloc}

\noi
There is a non-local generalization \ct{sts:83} of the solutions 
in \rf{loc} that reads as
\be
\begin{split}
R(e_{ij})~=~&\varepsilon(j\!-\!i)~e_{ij} 
+\delta_{i,j} \sum_{m\in\Z}\nu_{i,m}~e_{mm}  \\
P_{+}(e_{ij})~=~&\theta(j\!-\!i)~e_{ij}
+\frac{1}{2}\delta_{i,j} \sum_{m\in\Z}\nu_{i,m}~e_{mm} \\
P_{-}(e_{ij})~=~&\theta(i\!-\!j)~e_{ij}
-\frac{1}{2}\delta_{i,j} \sum_{m\in\Z}\nu_{i,m}~e_{mm}~.
\end{split} 
\lab{nloc}
\ee
The diagonal case $\nu_{i,j}=\nu_i  \delta_{i,j}$ corresponds to the previous
local solution. We claim that the non-local $R$-matrix possesses vanishing 
chiral Nijenhuis tensors (cf.~\eq{npmwedge}). To prove this, it is enough to 
consider $N_{\pm}(e_{ij}, e_{kl})$ for two basis elements $e_{ij}$ and
$e_{kl}$. $P_{+}$ and $P_{-}$ ``project'' onto links with weakly positive 
and weakly negative link length, respectively. The Lie-bracket preserves 
the total link length. So to give a non-zero contribution to 
$N_{\pm}(e_{ij}, e_{kl})$ {\em both} entries $e_{ij}$ and $e_{kl}$ have to 
be zero-length links. But the zero-length links are nothing but the diagonal
matrices and those commute trivially.

\noi 
In the non-local case, the sub-algebras ${\rm Im}(P_{+})$ and 
${\rm Im}(P_{-})$ form a direct sum if and only if the matrix $\nu_{i,j}$ 
is an involution:
\be
 {\rm Im}(P_{+}) \cap {\rm Im}(P_{-}) ~\subseteq~ \{0\} 
~~~~~~\Leftrightarrow~~~~~~  \nu^2~=~{\bf 1}~.
\lab{nldirectsum}
\ee

\subsection{\sf The $R$-Bracket}
\label{subsection:rliebracket}

\noi
{}For the local and non-local solutions \rf{loc} and \rf{nloc}, the 
$R$-matrix $R=R^{(0)}+R^{(1)}$ is a linear function of $\nu$, where the 
superscript ``$(0)$'' and ``$(1)$'' refer to the power of $\nu$. The 
$R$-bracket $[\,\cdot\, ,\,\cdot \, ]_R$ inherits this linear 
$\nu$-dependence, and can be split accordingly 
\be
 [\,\cdot\, ,\,\cdot\, ]_R ~=~
 [\,\cdot\, ,\,\cdot\, ]^{(0)}_R + [\,\cdot\, ,\,\cdot\, ]^{(1)}_R
\ee  
into two mutually compatible Lie-brackets $[\,\cdot\, ,\,\cdot\, ]^{(0)}_R$ 
and $[\,\cdot\, ,\,\cdot\, ]^{(1)}_R$. By definition, the $R$-bracket is a 
Lie pencil in $\nu$. We have
\be
\begin{split}
 [e_{ij},e_{kl}]_{R} ~=~& 
\frac{1}{2}\left[{\cal E}_{\nu}(j,i)+{\cal E}_{\nu}(l,k)\right]
\left(\delta_{j,k}~e_{il}-\delta_{i,l}~e_{kj}\right)  \\
=~&\left[\Theta_{\nu}(j,i)~\Theta_{\nu}(l,k)-
\Theta_{-\nu}(i,j)~\Theta_{-\nu}(k,l) \right] 
\left(\delta_{j,k}~e_{il}-\delta_{i,l}~e_{kj}\right)
\end{split}
\lab{nubracket}
\ee
for the local $R$-matrix \rf{loc}. This is a ${\rm gl}(\infty)$ Lie algebra 
$[ e_{ij} ,e_{kl} ] = \delta_{j,k} e_{il}-\delta_{i,l} e_{kj}$ with a 
$2$-cocycle-like prefactor. {}For the non-local $R$-matrix \rf{nloc}, only 
the first-order contribution in $\nu$ is changed. It reads
\be
 [e_{ij},e_{kl}]^{(1)}_{R} 
~=~\frac{1}{2}\delta_{i,j}~(\nu_{j,k}-\nu_{i,l})~e_{kl}
+\frac{1}{2}\delta_{k,l}~(\nu_{k,j}-\nu_{l,i})~e_{ij}~. 
\lab{nubracket1n}
\ee
It is a curious fact that links $e_{mm}$ of zero-length can never be produced
in a $R$-bracket $[\,\cdot\,, \,\cdot\, ]_R$ of the local or non-local type
(cf.\ \eqs{loc}{nloc}). Specifically, 
\be
  {\rm tr}\left( e_{mm} [u,v]_{R} \right)~=~0~.
\lab{curiouszerofact}
\ee
In contrast, the same does not hold for the standard ${\rm gl(\infty)}$
Lie-bracket $[\,\cdot\,, \,\cdot\, ]$, where for instance 
$[e_{ij},e_{ji}]=e_{ii}-e_{jj}$ yields two zero-length links if $i\neq j$.

\subsection{\sf Equations of Motion and Time Evolution}
\label{subsection:te}

\noi
It is interesting to write out the equations of motion in coordinates.
If we insert the local solutions \rf{loc} for $R$, $P_{+}$ and $P_{-}$ into 
the Lax \eq{ueom}, we get
\begin{subequations}
\br
\frac{\partial u_{ij}}{\partial t_n} 
&=&\frac{1}{2} \sum_{k\in\Z}  \left[{\cal E}_{\nu}(k,i)~(u^n)_{ik}~u_{kj}
- u_{ik}~{\cal E}_{\nu}(j,k)~(u^n)_{kj}  \right] \lab{rtime0} \\
&=& \sum_{k\in\Z}  \left[\Theta_{\nu}(k,i)~(u^n)_{ik}~u_{kj} 
- u_{ik}~\Theta_{\nu}(j,k)~(u^n)_{kj}  \right] \lab{ptime0} \\
&=& \sum_{k\in\Z}  \left[u _{ik}~\Theta_{-\nu}(k,j)~(u^n)_{kj}
- \Theta_{-\nu}(i,k)~(u^n)_{ik}~u_{kj}  \right]~,\lab{mtime0}
\er
\end{subequations}
respectively. The non-local generalization of \eq{rtime0} reads
\be
\begin{split}
\frac{\partial u_{ij}}{\partial t_n} 
~=&~\frac{1}{2} \sum_{k\in\Z} \left[ \varepsilon(k\!-\!i)~(u^n)_{ik}~u_{kj}
- u_{ik}~\varepsilon(j\!-\!k)~(u^n)_{kj} \right]  \\
&~+\frac{1}{2} u_{ij}\sum_{k\in\Z}(\nu_{k,i}-\nu_{k,j}) (u^n)_{kk}~.
\end{split} \lab{time0n}
\ee
The non-local generalizations of \eqs{ptime0}{mtime0} are similar.

\subsection{\sf Lattice Truncations}
\label{subsection:trunc}

\noi
We next address the question whether we can constrain the dynamical fields 
$u_{k}(x)$ without violating the equations of motion? As is easily seen in 
the matrix formalism, it is consistent with the equations of motion 
\rf{time0n} to deploy ``rectangular'' type of truncations of the matrix 
algebra ${\cal A}$ to a sub-algebra 
\be
{\cal A}_I \,:= \, 
\{ u=(u_{ij}) \in {\cal A} ~|~ u_{ij}\neq 0 ~\Rightarrow~ i,j \in I \}~
\ee
for some index-set $I \subseteq \Z$. {}From a difference operator
perspective, it is natural to consider  ``diagonal'' type of truncations.
Here, we discuss two  ``diagonal'' type of truncations that are often used 
in applications:
\begin{itemize}
\item
\underline{Truncation of the link length from below:}
It is consistent with the equations of motion to consider a truncated model 
with
\be
\forall k<N :~~ u_k (x) ~\equiv~ 0~.
\ee
To prove this, we note that the first (second) term on the rhs.\ of 
\rf{ptime0} has an index variable $k \ge i$ ($k \le j$), so that the left 
hand side $\partial u_{ij} / \partial t_n$ depends at least linearly on a 
link $u_{kj}$ ($u_{ik}$) with a signed length $j-k \le j-i$ ($k-i \le j-i$).
So if the field $u_{ij}$ (and its fellow fields with less or equal link 
length) are annihilated at some time $(t_1,t_2,\ldots)$, the equations of 
motions \rf{ptime0} cannot undo that for other times.
\item
\underline{Truncation of the link length from above:} A similar examination 
of \eq{mtime0} shows that there is also a consistent truncation from above:
\be
\forall k>M :~~ u_k  (x)~ \equiv ~0 \, .
\ee
\end{itemize}
The two truncation schemes are also consistent with the non-local solution 
\rf{time0n}, because when considering the left hand side 
$\partial u_{ij} / \partial t_n$, the non-local terms are always hidden 
behind at least one power of $u_{ij}$. By invoking both of the above 
truncation schemes, we get models with only a finite number of different 
fields  $u_M (x), \ldots, u_N (x)$ with link lengths between $M$ and $N$; 
all located inside an infinite universal enveloping construction. This fact 
renders the discrete hierarchy highly accessible for applications.

\section{\sf Poisson Brackets}
\label{section:pbrackets}

\noi
Before we proceed with constructing Poisson brackets, we need to introduce 
a few standard notions to fix the notation. A non-degenerate bilinear form 
\be
\langle u,v \rangle\,=\,{\rm tr}(uv)~=~\langle v,u \rangle \,
\ee
is inherited from the matrix trace (${\rm tr}$). Note that a {\em bi}-linear 
form $\langle \cdot, \cdot \rangle$, in contrast to a {\em sesqui}-linear 
form, has no internal transposition (or Hermitian conjugate for that matter). 
This is mainly to ensure that the bilinear form is invariant/associative:
\be
 \langle u\, , \, [v,w] \rangle~=~\langle [u,v]\, , \, w \rangle~.
\ee
The non-degenerate bilinear form gives rise to an identification of the 
algebra ${\cal A}$ with the set ${\cal A}^*$ of linear functionals on 
${\cal A}$. {}For a linear operator $R:{\cal A} \to {\cal A}$, the dual 
operator $R^*:{\cal A}^* \to {\cal A}^*$  becomes identified with the 
transposed operator
\be  
\langle v\, , \, R(u) \rangle~=~\langle R^{*}(v)\, , \, u \rangle~.
\ee
One may always decompose an operator $R$ in symmetric and skew-symmetric parts
\be
R_{\pm}~:=~\frac{R\pm R^*}{2}~.
\ee
Let us note for later that the non-local $R$-solution \rf{nloc} from
Section~\ref{subsection:nloc} decomposes as
\br
R_{-}(e_{ij})&=&\varepsilon(j\!-\!i)~e_{ij} 
+\frac{1}{2}\delta_{i,j} \sum_{m\in\Z}\nu_{[i,m]} e_{mm}~ \lab{Rnlocm} \\
R_{+}(e_{ij})&=& \frac{1}{2}\delta_{i,j} 
\sum_{m\in\Z}\nu_{\{i,m\}} e_{mm}~. \lab{Rnlocp}
\er
Notice that the skewsymmetric part $R_{-}$ in this case is again a $R$-matrix,
with vanishing chiral Nijenhuis tensors (cf.~\eq{npmwedge}). 

\noi
The adjoint action  ${\rm ad}(u):{\cal A} \to {\cal A}$ is defined as 
${\rm ad}(u)v:=[u,v]$. Because of the invariant/associative property of the
bilinear form, the coadjoint action (from right) 
${\rm ad}^*(u):{\cal A}^*\to {\cal A}^*$ is identified with {\em minus} 
the adjoint action
\be 
\langle {\rm ad}^*(u)v , w \rangle 
~=~ \langle v ,{\rm ad}(u)w \rangle 
~=~\langle v, [u,w] \rangle
~=~-\langle [u, v],w \rangle
~=~-\langle {\rm ad}(u) v ,w \rangle~.
\ee
The gradient $\nabla f$ of a function $f=f(u)$ on the dual space 
$u\in{\cal A}^*$ can be defined implicitly via the infinitesimal 
variational formula 
\be
 \delta f~=~\langle \delta u \, , \, \nabla f \rangle~. \lab{varfor}
\ee
Explicitly, the gradient is
\be
 \nabla \,=\,\sum_{i,j\in \Z} e_{ij} \frac{\partial }{\partial u_{ji}}~.
\ee
Notice the $i\!\leftrightarrow\!j$ transposition of indices in the above 
formula.

\subsection{\sf Conserved Charges}
\label{subsection:cc}

\noi
As is well-known, a hallmark of an integrable system is an infinity of 
conserved charges. In the discrete hierarchy, the charge densities are 
defined as 
\be
h_n(x)~=~\frac{1}{n} (L^n)_{0} \, (x)~=~
\frac{1}{n} (u^n)_{x,x}~,~~~~~~~~~~~~~~~~n > 0~,
\ee
and the charges themselves are defined as
\be
H_n~=~\sum_{x\in\Z} h_n(x)~=~\frac{1}{n} {\rm tr}(u^n)
~,~~~~~~~~~~~~~~~~~~~n > 0~,
\lab{hncharges}
\ee
where the trace in the last equation can be thought of as the sum of all 
closed loops build out of $n$ consecutive links. It follows directly from 
the Lax \eq{ueom} and from the cyclicity of the matrix trace 
${\rm tr}[u,v]\!=\!0$, that the charges $H_n$, $n > 0$, are conserved 
in time. Also, one observes easily that 
\be
 (L^n)_{k}\,(x)~=~(u^n)_{x,x+k}
~=~\frac{\partial H_{n+1}}{\partial u_{x+k,x}}~,
\lab{lnhn}
\ee
or, equivalently, 
\be
 \nabla  H_{n+1} ~=~ u^n \, .
\ee
The above facts are slightly more elaborate to establish purely within the 
difference operator approach (cf.\ for instance Theorem~III.2.7 
of \Ref{asterisque}). 

\noi
One may rewrite \ct{sts:83,harnad:93} the Lax \eq{ueom} as a standard
Hamiltonian equation on the dual space ${\cal A}^*$ 
\be
\frac{\partial u}{\partial t_n} ~=~\frac{1}{2}[R\nabla H_{n+1},u] 
~=~ -\frac{1}{2} {\rm ad}^*(R\nabla H_{n+1})u~,~~~~~~~~~~~~~u\in{\cal A}^*~.
\ee
It is useful to seek for Poisson bracket structures $\{ \cdot, \cdot \}_p$, 
$p=1, 2, \ldots, n+1$, such that the time evolutions can be reproduced as 
Hamiltonian flows with the conserved charges $H_{n+2-p}$ acting as generators
\be
\{  f(u), H_{n+2-p} \}_p 
~\equi{?}~\frac{\partial f}{\partial t_n} 
~\equi{\rf{varfor}}~
\langle \frac{\partial u}{\partial t_n}\, , \,\nabla f \rangle 
~\equi{\rf{ueom}}~\frac{1}{2} \langle [R(u^n),u] \, , \, \nabla f \rangle~.
\lab{rtimep}
\ee
In other words, find Poisson brackets such that the Lax \eq{ueom} can
be written as the Hamiltonian \eq{rtimep}. This will ensure the 
Lenard relations
\be
\{ \, \cdot \, , H_{n+1} \}_p~=~\{ \, \cdot \, , H_n \}_{p+1}
~,~~~~~~~~~~~~~~~~~~~~~~~~~~~n,p>0~.
\ee
Note that the $H_n$'s  play a double role as both Hamiltonian generators 
and as conserved quantities. This forces them to mutually ``commute'' 
in a Poisson sense: 
\be
\{ H_n, H_m \}_p~=~0~, ~~~~~~~~~~~~~~~~~~~~~~~~~~~n,m,p>0~.
\ee
{}For the local and the non-local solution discussed in 
Sections~\ref{subsection:loc}-\ref{subsection:nloc}, 
each Poisson bracket $\{ \cdot, \cdot \}_p$ will be a linear function of 
$\nu$-parameter
\be
\{ \cdot, \cdot  \}_p ~=~ \{ \cdot, \cdot  \}^{(0)}_p 
+ \{ \cdot, \cdot  \}^{(1)}_p
\ee  
of two mutually compatible Poisson brackets $\{ \cdot, \cdot \}^{(0)}_p$ 
and $\{ \cdot, \cdot \}^{(1)}_p$. This is sometimes referred to as a 
{\em Poisson pencil} in $\nu$.

\subsection{\sf 1st Poisson Bracket}
\label{subsection:1pb}

\noi
We now derive the first Poisson bracket from the $R$-matrix formalism.
{}From \eq{rtimep} and the Lax \eq{ueom}, we get
\be
\begin{split}
\{ f, H_{n+1} \}_1
~&\equi{?} \, \frac{\partial f}{\partial t_n}\, = \,
\langle \frac{\partial u}{\partial t_n}, \nabla  f \rangle \\
&=~\frac{1}{2}\langle \left[  R(u^n), u  \right], \nabla f \rangle
+\frac{1}{2}\langle [ u^n, u  ],  R(\nabla  f) \rangle  \\
&=~\frac{1}{2}\langle  u, \left[ \nabla  f,  R(u^n)\right] \rangle
+\frac{1}{2}\langle  u, \left[  R(\nabla  f), u^n \right] \rangle  \\
&=~\langle  u, [ \nabla f,  u^n ]_R \rangle
\,=\,\langle  u, \left[ \nabla f,  \nabla H_{n+1} \right]_R \rangle ~,
\end{split}
\ee
where the second term (which is identically zero) was added to achieve the 
required skewsymmetry. We immediately recognize the first $R$-theoretic 
Poisson structure \ct{sts:83}
\be
\{f,g\}_{R1}~:=~\langle u, [\nabla  f, \nabla g ]_R \rangle~.
\lab{r1theor}
\ee
{}For the local $R$-matrix solution \rf{loc}, this leads to 
\be
\{  u_{ij} ,  u_{kl} \}_{R1} ~=~ \langle u, [e_{ji},e_{lk}]_R \rangle
~=~\frac{1}{2}\left[{\cal E}_{\nu}(i,j) +{\cal E}_{\nu}(k,l)\right]
\left(\delta_{i,l}~u_{kj}-\delta_{j,k}~u_{il}\right)~.
\lab{r1bracket}
\ee
The local bracket $\{ \cdot , \cdot \}_{R1}$ has a pictorial interpretation
as a concatenation of two ``incoming'' links with a $2$-cocycle-like 
prefactor. {}For a non-local $\nu$, the first-order contribution in $\nu$ is
\be
\{  u_{ij} ,  u_{kl} \}^{(1)}_{R1}  ~=~ 
\frac{1}{2}\delta_{i,j}~(\nu_{i,l}-\nu_{j,k})~u_{kl}
+\frac{1}{2}\delta_{k,l}~(\nu_{l,i}-\nu_{k,j})~u_{ij}~. 
\lab{r1bracket1n}
\ee
The bracket operation $\{ \cdot, \cdot \}^{(1)}_{R1}$ preserves at least one 
of the ``incoming'' links $u_{ij}$ and $u_{kl}$ (up to an overall factor). 
To have a non-vanishing ``outgoing'' link, say $u_{ij}$, the other 
``incoming'' link $u_{kl}$ should have zero link length, and it should 
``interact at a distance'' with an endpoint of the first  ``incoming''
link. The interaction at a distance is mediated through a non-vanishing, 
non-local matrix element $\nu_{n,m}$. In the difference operator language, 
the local bracket transforms into
\be
\begin{split}
 \{  u_{n}(x) ,  u_{m}(y) \}_{R1} 
~=~&\frac{1}{2}\left[\varepsilon(n)-\nu_x~\delta_{n,0}
+ \varepsilon(m)-\nu_y~\delta_{m,0}\right]  \\  
& \times \left[u_{n+m}(x)~\delta_{x+n,y}
-u_{n+m}(y)~\delta_{x,y+m}\right]~.
\end{split}
\lab{r1bracketdiff}
\ee

\subsection{\sf Higher Poisson Brackets}

\noi
We now generalize the method of Section~\ref{subsection:1pb} to the higher 
brackets. Postponing the question of the Jacobi identity, let us tentatively 
write down skewsymmetric candidates for the $(p+1)$th bracket structure 
$\{ \cdot , \cdot \}_{p+1,r}$, labeled by an integer $r=0,1,\ldots, p$. 
{}From the \eq{rtimep} and the Lax \eq{ueom}, we get
\be
\begin{split}
& \{ f, H_{n+1-p} \}_{p+1,r}\,\equi{?} \, \frac{\partial f}{\partial t_n}
\, =\, \langle \frac{\partial u}{\partial t_n}, \nabla f \rangle \\
&=~\frac{1}{2}\langle \left[  R(u^n), u  \right], \nabla  f \rangle
+\frac{1}{2}\langle [ u^{n-p}, u  ],  
R(u^r~(\nabla f)~ u^{p-r}) \rangle \\
&=~\frac{1}{2}\langle u,\left[ \nabla f, 
R(u^r~u^{n-p}~u^{p-r})\right] \rangle 
 +\frac{1}{2}\langle u,\left[R(u^r~(\nabla  f)~ 
u^{p-r}),u^{n-p}\right]\rangle  \\
&=~\frac{1}{2}\langle  u, \left[ \nabla  f,  
R(u^r~(\nabla  H_{n+1-p})~u^{p-r})\right] \rangle 
+\frac{1}{2}\langle  u, \left[  R(u^r~(\nabla f)~u^{p-r}), 
 \nabla H_{n+1-p} \right] \rangle~,
\end{split}
\lab{rtimephigh}
\ee
where the second term (which is identically zero) was added to achieve the
required skewsymmetry. In this way we obtain the $r$'th candidate for the 
$(p+1)$'th $R$-matrix Poisson structure
\be
\{f,g\}_{p+1,r}~:=~\frac{1}{2}\langle  u, \left[ \nabla  f,  
R(u^r~(\nabla g)~u^{p-r})\right] \rangle 
 +\frac{1}{2}\langle  u, \left[  R(u^r~(\nabla f)~u^{p-r}), 
\nabla g \right] \rangle~.
\ee
The \eq{rtimephigh} is an inhomogeneous linear equation in the Poisson 
structure, with source terms generated from the Lax \eq{ueom}. We may not 
have properly identified possible homogeneous Poisson bracket parts 
$\{ \cdot , \cdot \}_{p+1}^{(H)}$ that commute with all the charges $H_n$, 
$n>0$. Besides these homogeneous contributions, the potential bracket 
candidates are convex linear combinations of the basis brackets 
$\{\cdot ,\cdot \}_{p+1,r}$, $r=0,1,\ldots, p$. Again we stress that most of 
brackets are going to be discarded, as they will not satisfy the Jacobi 
identity.

\subsection{\sf 2nd Poisson Bracket}

\noi
As we saw in the last Section, the potential bracket candidates for a 
quadratic bracket include the convex linear combinations of the two basis 
elements $\{\cdot, \cdot \}_{2,0}$ and $\{\cdot, \cdot \}_{2,1}$. The 2nd 
$R$-theoretic Poisson structure \ct{oevel:89} turns out to be the symmetric 
average
\be
\{f,g\}_{R2} ~:=~ \frac{1}{2}\{f,g\}_{2,0}+\frac{1}{2}\{f,g\}_{2,1} \\
~=~ \frac{1}{4}\langle u,\left[\nabla f,R\{u,\nabla g\}_{+}\right]\rangle 
-(f \leftrightarrow g)~,
\lab{r2theor}
\ee
where $\{u,v\}_{+}:=uv+vu$. If the Jacobi identity holds, one may show
that $\{ \cdot , \cdot \}_{R2}$ is always compatible with the first 
$R$-theoretic bracket $\{ \cdot , \cdot \}_{R1}$. This follows by shifting
$u \to u +\lambda {\bf 1}$ in \eq{r2theor}, because the shifted 2nd Poisson 
bracket
\be
\{f,g\}_{R2} \, (u \!+\!\lambda  {\bf 1})
~=~\{f,g\}_{R2} \,(u) + \lambda  \, \{f,g\}_{R1} \,(u)
\ee
can be re-interpreted as a Poisson pencil between the two brackets. (It is
enough to let $f$ and $g$ be linear functions of $u$, so that $\nabla f$ 
and $\nabla g$ are $u$-independent, which simplifies the above argument.)
Moreover, one may prove \ct{oevel:89} that sufficient conditions for the 
Jacobi identity for the $\{ \cdot , \cdot \}_{R2}$ bracket are, that $R$ and
$R_{-}$ satisfy the modified Yang-Baxter equations ${\rm YB}_{\alpha}(R)=0$ 
and ${\rm YB}_{\alpha}(R_{-})=0$ with the same parameter $\alpha$. This indeed
is the case for the local and non-local solutions (cf. \eqs{loc}{nloc}).

\noi
A generalization of the quadratic bracket is due to Suris \ct{suris:93}.
He defines a bracket
\be
\begin{split}
2 \{f,g\}_{S2}~:=~& 
\langle A_1( (\nabla f)~u ), (\nabla g)~u \rangle
- \langle A_2( u \nabla f), u \nabla g \rangle    \\
&+\langle S_1(u \nabla f  ), (\nabla g)~u \rangle
- \langle S_2( (\nabla f)~u), u \nabla g \rangle~, 
\end{split}
\lab{suris2}
\ee
where $A_1$, $A_2$, $S_1$, and $S_2$ are linear maps ${\cal A} \to {\cal A}$
satisfying $A^*_i=-A_i$ and $S^*_1=S_2$. 
We {\em assume everywhere in Section~\ref{section:pbrackets}} that
\be
    R~=~A_1+S_1~=~A_2+S_2~,
\lab{ras}
\ee
and that both $A_1$ and $A_2$ satisfy the modified Yang-Baxter equation 
${\rm YB}_{\alpha}(A_i)=0$, $i=1,2$.

\noi
With the above assumptions one can show that the two Suris conditions
\be
\begin{split}
 & 2 S_1 [u,v]_{A_2}~=~[S_1(u),S_1(v)]  \\
 & 2 S_2 [u,v]_{A_1}~=~[S_2(u),S_2(v)] 
\end{split}
\lab{suriscond}
\ee
are sufficient for the Jacobi identity to hold. Also, they imply the modified
Yang-Baxter equation ${\rm YB}_{\alpha}(R)=0$, and that 
$\{\cdot ,\cdot \}_{R1}$ and $\{\cdot ,\cdot \}_{S2}$ are compatible Poisson
brackets.

\noi
Note that the opposite does {\em not} hold, \ie that ${\rm YB}_{\alpha}(R)=0$ 
does not necessarily imply the Suris conditions \rf{suriscond}. We give a 
counterexample in Appendix~\ref{section:app22}. Also, the 2nd $R$-theoretic
Poisson structure $\{\cdot ,\cdot \}_{R2}$ can be seen as a special case of
the Suris construction if one let $A_1=A_2=R_{-}$ and $S_1=S_2=R_{+}$,
because in this case the Suris condition 
$2R_{+}[u,v]_{R_{-}}=[R_{+}(u),R_{+}(v)]$ {\em does} follow from the modified
Yang-Baxter equations ${\rm YB}_{\alpha}(R)=0$ and
${\rm YB}_{\alpha}(R_{-})=0$.

\noi
It is known that a compatible quadratic Poisson structure for the discrete
hierarchy is not unique \ct{suris:93}, although a full classification of 
ambiguities is still an open problem. Here, we give a family of solutions 
that can be described using the quadratic Suris bracket. To this end, define 
a skewsymmetric linear map $\Omega:{\cal A} \to {\cal A}$
\be 
\Omega~:=~\frac{A_1-A_2}{2}~=~\frac{S_2-S_1}{2}~=~-\Omega^*~,
\ee
where we used \eq{ras} in the second equality. One may decompose the Suris
variables entirely in terms of $R$ and $\Omega$:
\be
 A_1~=~R_{-}+\Omega~,~~~~A_2~=~R_{-}-\Omega~,~~~~
S_1~=~R_{+}-\Omega~,~~~~S_2~=~R_{+}+\Omega~,
\ee
as well as the Suris bracket itself
\be
  \{\cdot , \cdot \}_{S2}~=~ 
\{\cdot ,\cdot \}_{R2} + \{\cdot ,\cdot \}_{\Omega 2}~,
\ee
where 
\be
\{f,g\}_{\Omega 2}~:=~ 
\frac{1}{2}\langle\Omega \left[u,\nabla f\right],
\left[u,\nabla g\right]\rangle~.
\ee
The structure $\{\cdot ,\cdot \}_{\Omega 2}$ is {\em not} required to satisfy
the Jacobi identity, and hence it is not necessarily a Poisson bracket,
although this turns out to be the case for our example below. A sufficient
condition for this to happen is given by the Yang-Baxter equation 
${\rm YB}_{0}(\Omega)=0$. In general, the structure 
$\{\cdot ,\cdot \}_{\Omega 2}$ does not contribute to the Hamiltonian 
\eq{rtimep}, because $\{H_n , \, \cdot \, \}_{\Omega 2}=0$, $n>0$. Hence, 
even for the more general Suris bracket $\{\cdot ,\cdot \}_{S2}$, the 
dynamics is governed by the $\{\cdot ,\cdot \}_{R2}$ bracket alone.

\noi
We claim that the non-local $R$-solution \rf{nloc}, together with the choice
\be
\Omega(e_{ij})~=~\delta_{i,j} \sum_{m\in\Z} \omega_{i,m} e_{mm}~,
\lab{omeganloc}
\ee
for some skewsymmetric matrix $\omega_{i,j}=-\omega_{j,i}$, meets all the 
conditions of the Suris construction. The proofs are very similar to the 
discussion given in Section~\ref{subsection:nloc}. First of all, both $A_1$ 
and $A_2$ are of the same form as the non-local $R$-solution \rf{nloc}, 
and therefore they too have vanishing chiral Nijenhuis tensors 
$N_{\pm}(A_i)=0$, and hence ${\rm YB}_{1}(A_i)=0$. Secondly, both sides of 
the Suris conditions \rf{suriscond} vanish. {}For instance, the lhs.\ is of 
the form $S_i(w)$, where $w=2 [u,v]_{A_j}$. Because of the special form of 
the two $S_{i}$ maps, $i=1,2$, only diagonal parts of $w$ could potentially 
contribute. On the other hand, diagonal parts of $[u,v]_{A_j}$ do not exist 
according to \eq{curiouszerofact}. So the lhs.\ is zero. The rhs.\ is zero, 
because both $S_i(u)$ and $S_i(v)$ are diagonal matrices, and hence commute.

\noi
Let us write down the Suris quadratic bracket 
$\{\cdot ,\cdot \}_{S2}
=\{\cdot ,\cdot \}^{(0)}_{S2}
+\{\cdot ,\cdot \}^{(1)}_{S2}$ in detail
\begin{subequations}
\br
 \{ u_{ij},u_{kl} \}^{(0)}_{S2}&=&\frac{1}{2} \left[\varepsilon(k\!-\!i)
+\varepsilon(l\!-\!j)\right]  u_{il}~u_{kj}~,\lab{s2bracket0} \\
 \{ u_{ij},u_{kl} \}^{(1)}_{S2}&=&\omega_{ij,kl}~ u_{ij} ~u_{kl}~,
\lab{s2bracket1a}
\er
\end{subequations}
where 
\be
 \omega_{ij,kl}~:=~\frac{1}{4} 
(\nu_{\{i,l\}} -\nu_{\{j,k\}}+\nu_{[k,i]} +\nu_{[j,l]})
~+~\frac{1}{2}(\omega_{j,l}+\omega_{l,i}+\omega_{i,k}+\omega_{k,j}) 
~=~ -\omega_{kl,ij}~.
\lab{s2bracket1n} 
\ee 
{}For simplicity, we have collected all the $\omega_{i,j}$-terms inside the 
$\{ \cdot ,\cdot \}^{(1)}_2$-part. (Strictly speaking, this represents 
a minor abuse of notation, because $\omega_{i,j}$ does not need to be 
first order in $\nu$.) {}Figuratively speaking, the Suris bracket 
$\{ \cdot ,\cdot \}_{S2}$ consists of two parts 
$\{ \cdot ,\cdot \}^{(1)}_{S2}$ and $\{ \cdot ,\cdot \}^{(0)}_{S2}$ 
that represent elastic and inelastic scattering of two ``incoming'' links,
respectively. In other words, the first order bracket 
$\{ \cdot ,\cdot \}^{(1)}_{S2}$ preserves the two ``incoming'' links 
$u_{ij}$ and $u_{kl}$, while the two ``incoming'' links exchange a pair of 
endpoints in the zero order bracket $\{ \cdot ,\cdot \}^{(0)}_{S2}$. 

\noi
If we restrict the $\Omega$-contribution to be the form 
$\omega_{i,j}=\omega_{i-j}=-\omega_{j-i}$, the quadratic bracket reads
\be
\begin{split}
 \{  u_{n}(x) ,  u_{m}(y) \}^{(0)}_{S2} 
~=~&\frac{1}{2}\sum_{k\in\Z} \left[\varepsilon(k)
+\varepsilon(k\!+\!m\!-\!n)\right]
u_{n-k}(y)~u_{m+k}(x)~\delta_{x+k,y} \\
 \{  u_{n}(x) ,  u_{m}(y) \}^{(1)}_{S2} 
~=~& \frac{1}{2}u_{n}(x)~u_{m}(y)\sum_{k\in\Z} [
\nu_{x}~\delta_{k,-m}-\nu_{y}~\delta_{k,n} \\
&-\omega_{k+m-n}+\omega_{k+m}
-\omega_{k}+\omega_{k-n} ]~\delta_{x+k,y}
\end{split}
\lab{s2bracketdiff}
\ee
in the difference operator language.

\subsection{\sf 3rd Poisson Bracket}

\noi
The potential bracket candidates for a cubic Poisson bracket are convex 
linear combinations of the three basis elements $\{\cdot, \cdot \}_{3,0}$, 
$\{\cdot, \cdot \}_{3,1}$ and $\{\cdot, \cdot \}_{3,2}$. The 3rd $R$-theoretic
bracket \ct{oevel:89,li:89} turns out to be given entirely by the candidate 
$\{\cdot,\cdot\}_{3,1}$:
\be
\{f,g\}_{R3}~:=~\{f,g\}_{3,1}
~=~\frac{1}{2}\langle  u, \left[ \nabla  f, R(u~(\nabla  g)~u)\right] \rangle 
 +\frac{1}{2}\langle u, \left[  R(u~(\nabla  f)~u), \nabla  g \right] \rangle~,
\lab{r3theor}
\ee
One may show \ct{oevel:89,li:89} that the three $R$-theoretic brackets
$\{\cdot ,\cdot \}_{R1}$, $\{\cdot ,\cdot \}_{R2}$ and $\{\cdot ,\cdot \}_{R3}$
are compatible Poisson structures if both $R$ and $R_{-}$ satisfy the 
modified Yang-Baxter equation ${\rm YB}_{\alpha}(R)=0$ and 
${\rm YB}_{\alpha}(R_{-})=0$ with the same parameter $\alpha$. In general, 
the third bracket $\{\cdot , \cdot\}_{R3}$ is {\em not} compatible with the
Suris quadratic bracket $\{\cdot , \cdot\}_{S2}$.

\noi
We derive
\be
\begin{split}
\{  u_{ij} ,  u_{kl} \}_{R3}
~=~& \frac{1}{2} \sum_m[{\cal E}_{\nu}(m,i)
+{\cal E}_{\nu}(l,m)]~ u_{il}~u_{km}~u_{mj}  \\
&-\frac{1}{2} \sum_m [{\cal E}_{\nu}(j,m)
+{\cal E}_{\nu}(m,k)]~u_{im}~u_{ml}~u_{kj}~
\end{split}
\lab{kuper31bracket}
\ee
for the non-local $R$-solution \rf{nloc}. Its local first order $\nu$ terms 
are
\be
\{  u_{ij} ,  u_{kl} \}^{(1)}_{R3}
~=~\frac{1}{2} u_{ij}~[ u_{ki}~\nu_i~u_{il}
-u_{kj}~\nu_j~u_{jl}]
~-~[(i,j) \leftrightarrow (k,l)]~,
\lab{kuper31bracket1}
\ee
while the non-local first order $\nu$ terms read
\be
\{  u_{ij} ,  u_{kl} \}^{(1)}_{R3}
~=~\frac{1}{2} u_{ij} \sum_{m\in\Z} u_{km}~(\nu_{m,i}-\nu_{m,j})~u_{ml}
~-~[(i,j) \leftrightarrow (k,l)]~.
\lab{kuper31bracket1n}
\ee
There is an interesting duality between the $1$st and the $3$rd bracket, which 
(formally) facilitates the proof of the Jacobi identity for the third 
bracket. {}Following Oevel and Ragnisco \ct{oevel:89}, one notices that 
matrix inversion $u \mapsto u^{-1}$ maps the first and third bracket into 
each other up to an overall minus sign. In detail, consider linear functionals 
$f(u)=\langle a, u \rangle$ and $g(u)=\langle b, u \rangle$ for some constant 
algebra elements $a,b \in {\cal A}$. Then for invertible $u$'s
\br
\nabla f&=&a~,~~~~~~~~~~~~~~~~~~~~~~~~~~~~~~~~~
\nabla g~=~b~, \\
\nabla (f(u^{-1}))&=&-u^{-1}au^{-1}
~~~~~~~{\rm and}~~~~~~~
\nabla (g(u^{-1}))~=~-u^{-1}bu^{-1}~.
\er
It follows from the definitions \rfs{r1theor}{r3theor} that
\be
\begin{split}
\{ f(u^{-1}),g(u^{-1}) \}_{R3}\,(u)
&=~\frac{1}{2}\langle  \left[ u, u^{-1}au^{-1} \right], R(b) \rangle 
~-~(a \leftrightarrow b) \\
&=~\frac{1}{2}\langle  \left[ a, u^{-1} \right], R(b) \rangle 
~-~(a \leftrightarrow b) \\
&=~-\{ f,g \}_{R1}\,(u^{-1})~.
\end{split}
\ee
This provides a proof of the Jacobi identity within the matrix group of 
invertible matrices (cf.\ footnote \ref{footnotalginv}). Similarly, one 
notices that the $2$nd bracket is self-dual under $u \mapsto u^{-1}$ up 
to an overall minus sign
\be
\begin{split}
\{ f(u^{-1}),g(u^{-1}) \}_{R2}\, (u)
&=~\frac{1}{2}\langle \left[ u, u^{-1}au^{-1} \right],
 R\{u,u^{-1}bu^{-1}\}_{+} \rangle ~-~(a \leftrightarrow b) \\
&=~\frac{1}{2}\langle  \left[ a, u^{-1} \right], 
R\{b,u^{-1}\}_{+}  \rangle ~-~(a \leftrightarrow b) \\
&=~-\{ f,g \}_{R2}\, (u^{-1})~.
\end{split}
\ee
Interestingly, the structure $\{\cdot ,\cdot \}_{\Omega 2}$ is also self-dual
under $u \mapsto u^{-1}$, but with an overall plus sign
\be
\begin{split}
\{ f(u^{-1}),g(u^{-1}) \}_{\Omega 2}\, (u)
&=~\frac{1}{2}\langle \Omega \left[ u, u^{-1}au^{-1} \right],
 \left[u,u^{-1}bu^{-1}\right] \rangle ~-~(a \leftrightarrow b) \\
&=~\frac{1}{2}\langle \Omega \left[ a, u^{-1} \right],
\left[b,u^{-1}\right]  \rangle ~-~(a \leftrightarrow b) \\
&=~+\{ f,g \}_{\Omega 2}\, (u^{-1})~,
\end{split}
\ee
so different parts of the Suris bracket $\{\cdot ,\cdot \}_{S2}$ has
different transformation properties under duality. Of course, one may claim 
the Suris bracket is self-dual under $u \mapsto u^{-1}$ if one simultaneously 
changes the sign of $\Omega$, or equivalently, one simultaneously exchanges 
$A_1\leftrightarrow A_2$ and $S_1\leftrightarrow S_2$.

\noi
{\sc Acknowledgment:}~This work has been partially supported by DOE grant 
DOE-ER-40173.

\vspace{1cm}
\appendix

\section{\sf The Kupershmidt Bi-Poisson Structure}
\label{section:appkuper}

\noi
Here we translate our results into the $q_k(x)$ fields used by Kupershmidt 
\ct{asterisque}. He uses an anti-normal ordered Lax operator of the form
\be
L~=~\sum_{k \in \Z} \Delta^{-k} q_k(x)~. 
\ee
Comparing with \eq{ourlax}, one derives the translation formula
\be 
q_{k}(x)~=~u_{k+x,x}~=~u_{-k}(k\!+\!x)~.
\ee
We may facilitates the $u_k(x) \leftrightarrow q_k(x)$ translation of the 
Poisson structures by the following observation. {}First of all, from the 
matrix definition $u_{k}(x)=u_{x,x+k}$, one notices that a change of 
variables $u_k(x) \leftrightarrow q_k(x)$ corresponds to a transposition of
the link matrix $u_{ij} \leftrightarrow u_{ji}$. Secondly, notice that the
$\{\cdot ,\cdot \}_{R1}$ bracket \rfs{r1bracket}{r1bracket1n} and the 
$\{\cdot ,\cdot \}_{S2}$ bracket \rfs{s2bracket0}{s2bracket1a} are invariant 
under a transposition $u \to u^T$ of the $u$-matrix combined with a change 
of the sign $\nu\to -\nu$ of the  $\nu$-matrix. Hence the 
$u_k(x) \leftrightarrow q_k(x)$ translation simply amounts to a change of 
the sign of $\nu$. The $\{\cdot ,\cdot \}_{R1}$ bracket \rf{r1bracketdiff} 
and the $\{\cdot ,\cdot \}_{S2}$ bracket \rf{s2bracketdiff} become
\be
\begin{split}
 \{ q_{n}(x) , q_{m}(y) \}_{R1} ~=~&
\frac{1}{2}\left[\varepsilon(n)+\nu_x~\delta_{n,0}
+ \varepsilon(m)+\nu_y~\delta_{m,0}\right] \\
& \times \left[q_{n+m}(x)~\delta_{x+n,y}
-q_{n+m}(y)~\delta_{x,y+m}\right]~,\\
 \{ q_{n}(x) , q_{m}(y) \}^{(0)}_{S2} 
~=~&\frac{1}{2}\sum_{k\in\Z} \left[\varepsilon(k)
+\varepsilon(k\!+\!m\!-\!n)\right]
q_{n-k}(y)~q_{m+k}(x)~\delta_{x+k,y}~, \\
 \{ q_{n}(x) , q_{m}(y) \}^{(1)}_{S2} 
~=~& \frac{1}{2}q_{n}(x)~q_{m}(y)\sum_{k\in\Z} [
-\nu_{x}~\delta_{k,-m}+\nu_{y}~\delta_{k,n} \\
&-\omega_{k+m-n}+\omega_{k+m}
-\omega_{k}+\omega_{k-n} ]~\delta_{x+k,y}~,~~~~~~~~~~~~~n,m\in\Z~.
\end{split}
\lab{qbracketdiff}
\ee
In the original model of \Ref{asterisque}, the fields corresponding to 
positive link lengths are constrained 
\be
q_{-1} (x)~\simeq~ 1 \quad {\rm  and}   \quad  
\forall k\le -2:~ q_{k}(x)~\simeq~0~,
\ee
so that the Lax operator reads
\be
L~=~\Delta+\sum_{k \ge 0} \Delta^{-k} q_k(x)~. 
\ee
The constraints have to be consistent with the equations of motion, written 
either as Hamiltonian equations \rf{rtimep} or as Lax equations \rf{ueom} -- 
with or without use of Poisson brackets, respectively. Previously in 
Section~\ref{subsection:trunc}, we saw that the constraints
$\forall k\le -2: q_{k}(x)\simeq 0$ are consistent with the Lax formulation. 
Also, it is easy to check from the bi-Poisson structure \eq{qbracketdiff} 
that the constraints $\forall k\le -2: q_{k}(x)\simeq 0$ decouple from the 
theory in the Hamiltonian sense, \ie that the Hamiltonian vectorfields 
$\{ q_{k}(x),\, \cdot \, \} \simeq 0$ vanish for both brackets when $k\le -2$.
On the other hand, the constraint $q_{-1}(x) \simeq 1$ induces non-trivial 
conditions on the model. {}From the Lax \eq{leom} using ${\cal P}_{-}$ and 
the expansion \eq{lexpand}, one derives
\be
 \frac{\partial  q_{-1}(x)}{\partial t_n}
~=~\frac{1\!-\!\nu(x)}{2} q_{-1}(x)
\left[(L^n)_0 \, (x)- (L^n)_0\,(x\!-\!1)\right]~,
\ee
so consistency requires $\nu=1$. Moreover, in the Hamiltonian formulation, 
where one imposes that the field $q_{-1}(x)\!\equiv \!1$  ``Poisson commutes''
with the other fields $q_{n}(x)$, $n \ge 0$, one is lead to the choice 
$\nu=1$ and $\omega_{k}= kc -\varepsilon(k)=-\omega_{-k}$ with some immaterial 
constant $c\in\C$. (Again we stress that the on-shell dynamics are not 
affected by the $\{\cdot ,\cdot \}_{\Omega2}$ contributions.) 
With this choice, the brackets \rf{qbracketdiff} simplify to
\be
\begin{split}
  \{  q_{n}(x) ,  q_{m}(y) \}_{K1} 
~=~& q_{n+m}(x)~\delta_{x+n,y}-q_{n+m}(y)~\delta_{x,y+m}~,\\
 \{ q_{n}(x) , q_{m}(y) \}^{(0)}_{K2} 
~=~&\frac{1}{2}\sum_{k\in\Z} \left[\varepsilon(k)
+\varepsilon(k\!+\!m\!-\!n)\right]
q_{n-k}(y)~q_{m+k}(x)~\delta_{x+k,y}~, \\
 \{ q_{n}(x) , q_{m}(y) \}^{(1)}_{K2} 
~=~& \frac{1}{2}q_{n}(x)~q_{m}(y)\sum_{k\in\Z} [
\varepsilon(k)+\varepsilon(k\!+\!m\!-\!n) \\
&-\varepsilon(k\!+\!m\!+\!\frac{1}{2})-
\varepsilon(k\!-\!n\!-\!\frac{1}{2})]~\delta_{x+k,y}
~,~~~~~~~~~~~~~n,m~\ge~ 0~,
\end{split}
\lab{kuperbracketdiff}
\ee
which agree with the formula (III.3.4) and the formula (III.4.15a-c) in 
\Ref{asterisque}.

\section{\sf Example: ${\rm Mat}_{2 \times 2}(\C)$ }
\label{section:app22}

\noi
Here we give a counterexample, that shows that 
${\rm YB}_{\alpha}(R)={\rm YB}_{\alpha}(A_1)={\rm YB}_{\alpha}(A_2)=0$, 
taken together with the relation $R=A_i+S_i$, does {\em not} necessarily imply 
the Suris conditions \rf{suriscond}.

\noi
Consider the $4$-dimensional associative algebra 
${\cal A}={\rm Mat}_{2 \times 2}(\C) \cong \C^4$. A convenient basis is given 
by the $2\times 2$ unit-matrix $\sigma_4\equiv{\bf 1}$ and the three Pauli 
$\sigma_i$ matrices, $i=1,2,3$, which satisfy the relation
\be
\sigma_i~\sigma_j~=~\delta_{ij}{\bf 1}+i \sum_{k=1}^3\epsilon_{ijk}\sigma_k
~,~~~~~~~~~~~~~~~~~~~~~i,j~=~1,2,3~,
\ee
where $\epsilon_{ijk}$ is the $3$-dimensional Levi-Civita symbol.
A non-degenerate, associative/invariant bilinear form is inherited from the 
matrix trace (${\rm tr}$):
\be
\langle\sigma_{\mu}  ,\sigma_{\nu} \rangle
~=~{\rm tr}\left(\sigma_{\mu}~ \sigma_{\nu}\right)~=~ 2 \delta_{\mu,\nu}
~,~~~~~~~~~~~~~~~~~\mu,\nu=1,2,3,4~.
\ee
We now search for $R$-matrix solutions to the modified Yang-Baxter 
equation ${\rm YB}_{\alpha}(R)=0$, where $\alpha\in\C$ is a given fixed
complex number. Let us consider a linear injective map $\Phi$
\be
\C^{\, 3}~\ni~ \vec{r}=(r_1,r_2,r_3)
~~~~~\stackrel{\Phi}{\mapsto}~~~~~R~ \in~ {\rm End}({\cal A})~,
\ee
that maps a complex rotation vector $\vec{r}$ into its rotation matrix $R$
\be
\begin{split}
R(\sigma_i)~:=~&i \sum_{j,k=1}^3 \epsilon_{ijk} r_j \sigma_k
~,~~~~~~~~~~~~~~~~~~~~~~~i~=~1,2,3~, \\
R({\bf 1})~:=~&0~.
\end{split}
\ee
In other words, $R$ rotates the basis of $\sigma_i$ matrices, $i=1,2,3$,
around the rotation axis $\vec{r}$. Note that $R=-R^*$ is skewsymmetric. 
The $R$-bracket reads
\be
\begin{split}
 [\sigma_i,\sigma_j]_R~=~& \sigma_{[i}~ r_{j]}
~,~~~~~~~~~~~~~~~~~~~~~~~~~~~~~i,j~=~1,2,3~, \\
{} [{\bf 1} , \cdot \, ]_R~=~&0~.
\end{split}
\ee
We claim that ${\rm YB}_{\vec{r} \cdot \vec{r}}(R)=0$, \ie that
\be
2R[\sigma_{\mu},\sigma_{\nu}]_R
~=~[R(\sigma_{\mu}),R(\sigma_{\nu})]
+\vec{r} \cdot \vec{r}~ 
[\sigma_{\mu},\sigma_{\nu}]~,~~~~~~~~~~~~~\mu,\nu=1,2,3,4~,
\lab{ybmunu}
\ee
where $\vec{r} \cdot \vec{r} :=\sum_{i=1}^3r_i^2$ is a ``bilinear''
length-square, \ie without a complex conjugation. Technically, since the 
$\sigma_4$-sector is trivial, the \eq{ybmunu} reduces to a  ``dual'' 
Yang-Baxter identity 
\be
\sum_{j,k=1}^3 \epsilon_{ijk} 
\left(\left[R(\sigma_j),R(\sigma_k)\right]-2R\left[\sigma_j,\sigma_k\right]_R
+ \vec{r} \cdot \vec{r}~ \left[\sigma_j,\sigma_k\right]  \right)~=~0
~,~~~~~~~~~i~=~1,2,3~,  
\ee
which is easy to verify by direct calculation.

\noi
Now let us apply this fact to a specific example. Define five rotation vectors 
\be
 \vec{r}~=~(1,i,\sqrt{\alpha})~,~~~~~~~~~~~~~~~~~~~~~
\vec{s}_1~=~(1,i,0)~=~-\vec{s_2}~. 
\ee
and
\be
 \vec{a}_1~=~\vec{r}-\vec{s}_1~=~(0,0,\sqrt{\alpha})~,~~~~~~~~~~~~~
\vec{a}_2~=~\vec{r}-\vec{s}_2~=~(2,2i,\sqrt{\alpha})~. 
\ee
The corresponding five rotation matrices are skewsymmetric
\be
\begin{split}
A_i~=~&\Phi(\vec{a}_i)~=~-A^*_i~,\\
S_1~=~&\Phi(\vec{s}_1)~=~\Phi(-\vec{s}_2)
~=~-\Phi(\vec{s}_2)~=~-S_2~=~S^*_2~,\\
R~=~&\Phi(\vec{r})~=~\Phi(\vec{a}_i+\vec{s}_i)
~=~\Phi(\vec{a}_i)+\Phi(\vec{s}_i)~=~A_i+S_i~,~~~~~i~=~1,2~,
\end{split}
\ee
and they each satisfy a (modified) Yang-Baxter equation
\be
 {\rm YB}_{\alpha}(R)~=~0~,~~~~~~~{\rm YB}_{\alpha}(A_i)~=~0~,~~~~~~~
{\rm YB}_{0}(S_i)~=~0~,~~~~~~~~~i~=~1,2~.
\lab{jambalaja}
\ee
Using ${\rm YB}_{0}(S_1)=0$, one may reduce the Suris operator 
\be
(u,v) ~~~\mapsto~~~ 2 S_1 [u,v]_{A_2}-[S_1(u),S_1(v)] ~=~ 2 S_1 [u,v]_R
\ee
to only one term (cf.\ \eq{suriscond}). Therefore, the ``dual'' Suris 
condition simplifies to
\be
\frac{1}{2}\sum_{j,k=1}^3 \epsilon_{ijk} S_1[\sigma_j,\sigma_k]_R
~=~i\sum_{j=1}^3 r_j s_{1,[j}~\sigma_{i]} 
~=~-i~ \vec{r} \cdot \vec{\sigma}~ s_{1,i} ~\neq~ 0~,~~~~~i~=~1,2~.
\ee
Thus the Suris conditions \rf{suriscond} are not met, despite \eq{jambalaja}.

\noi
It is however generally valid that the two Suris conditions \eq{suriscond}, 
the modified Yang-Baxter equation ${\rm YB}_{\alpha}(R)=0$, taken 
together with the relation $R=A_i+S_i$, imply the two modified Yang-Baxter 
equations ${\rm YB}_{\alpha}(A_i)=0$ for the skewsymmetric maps $A_i$,
$i=1,2$ (cf.\ Theorem 2 in \Ref{oevel:96}).

\end{document}